\documentclass[%
 reprint,
superscriptaddress,
 amsmath,amssymb,
 aps,
prx,
longbibliography
]{revtex4-2}

\usepackage{graphicx}
\usepackage{dcolumn}
\usepackage{bm}
\usepackage{silence}
\usepackage[normalem]{ulem}

\usepackage{tikz,xcolor}
\usepackage[hidelinks]{hyperref}
\renewcommand{\d}{\textrm{d}}
\renewcommand{\v}[1]{\bm{{#1}}}
\newcommand{\mean}[1]{\langle {#1} \rangle}
\newcommand{\ddt}[1]{\frac{\textrm{d}{#1}}{\textrm{d}t}}

\renewcommand{\sp}{\mathcal S_{\textrm{\tiny +}}}
\newcommand{\PP}{\mathcal{P}}

\definecolor{blue}{rgb}{0,0,1}

\definecolor{dgreen}{rgb}{0.0,0.5,0.0}

\definecolor{lightBlue}{rgb}{0.,0.5,0.5}

\definecolor{EikeColor}{rgb}{0.9,0.1,0}

\usepackage{orcidlink}

\usepackage{siunitx}
\usepackage{relsize}
\usepackage{algorithmic,algorithm}
\usepackage{todonotes}
\usepackage[utf8]{inputenc}

\begin{document}

\title{
Physics-inspired machine learning for power grid frequency modelling
}

\author{Johannes Kruse \orcidlink{0000-0002-3478-3379}}
\affiliation{Forschungszentrum J\"ulich, Institute for Energy and Climate Research (IEK-STE), 52428 J\"ulich, Germany}
\affiliation{Institute for Theoretical Physics, University of Cologne, K\"oln, 50937, Germany}

\author{Eike Cramer \orcidlink{0000-0002-6882-5469} }%
\affiliation{Forschungszentrum Jülich GmbH,
Institute of Energy and Climate Research,
Energy Systems Engineering (IEK-10),
Jülich 52428,
Germany}

\author{Benjamin Schäfer \orcidlink{0000-0003-1607-9748}}%
\affiliation{Institute for Automation and Applied Informatics, Karlsruhe Institute of Technology, 76344 Eggenstein-Leopoldshafen, Germany}

\author{Dirk Witthaut \orcidlink{0000-0002-3623-5341}}%
\email{d.witthaut@fz-juelich.de}
\affiliation{Forschungszentrum J\"ulich, Institute for Energy and Climate Research (IEK-STE), 52428 J\"ulich, Germany}
\affiliation{Institute for Theoretical Physics, University of Cologne, K\"oln, 50937, Germany}

\date{\today}

\begin{abstract}

The operation of power systems is affected by diverse technical, economic and social factors. Social behaviour determines load patterns, electricity markets regulate the generation and weather-dependent renewables introduce power fluctuations. Thus, power system dynamics must be regarded as a non-autonomous system whose parameters vary strongly with time. However, the external driving factors are usually only available on coarse scales and the actual dependencies of the dynamic system parameters are generally unknown. Here, we propose a physics-inspired machine learning model that bridges the gap between large-scale drivers and short-term dynamics of the power system. Integrating stochastic differential equations and artificial neural networks, we construct a probabilistic model of the power grid frequency dynamics in Continental Europe. Its probabilistic prediction outperforms the daily average profile, which is an important benchmark. Using the integrated model, we identify and explain the parameters of the dynamical system from the data, which reveals their strong time-dependence and their relation to external drivers such as wind power feed-in and fast generation ramps. Finally, we generate synthetic time series from the model, which successfully reproduce central characteristics of the grid frequency such as their heavy-tailed distribution. All in all, our work emphasises the importance of modelling power system dynamics as a stochastic non-autonomous system with both intrinsic dynamics and external drivers. 

\end{abstract}

\maketitle

\section{Introduction}

Mitigation of climate change requires a comprehensive transformation of our economy and lifestyle, in particular the way we generate and utilise electric power \cite{rockstrom2017roadmap,rogeljEnergySystemTransformations2015}. Power plants based on fossil fuels must be replaced by renewable sources such as wind and solar power, which are volatile and uncertain \cite{staffell2018increasing}. Various sectors are being integrated,  for instance through electric heatpumps \cite{orths2019flexibility}, introducing numerous new interdependencies and increasing system complexity. The electric power system is at the heart of this transformation. Hence, understanding risks and guaranteeing stability of the electric power system is critical amidst far-reaching challenges \cite{witthaut2021collective}.

Power system operation is determined by various technical, economic and social influences and perturbations. Power generation from renewable sources is essentially determined by the weather \cite{heideSeasonalOptimalMix2010, collinsImpactsInterannualWind2018}, while the dispatch of conventional power plants is determined on various electricity markets \cite{linElectricityMarketsTheories2017}. Moreover, the load depends on the decisions and actions of millions of consumers \cite{anvariDatadrivenLoadProfiles2022}. As the power grid does not store electric energy, generation and load must be balanced at all times. On long time scales of hours, this is achieved by trading on electricity markets \cite{han2022complexity}. On short time scales of seconds and minutes, several layers of control reserves balance the grid, e.g., to counteract unforeseen perturbations and forecasting errors \cite{machowskiPowerSystemDynamics2008}. The activation of these reserves is mainly controlled by the grid frequency, which directly monitors the power imbalance: A scarcity of generation leads to a drop of the frequency, which is easily monitored anywhere in the grid. The stability of this load-frequency control system is challenged by the energy transformation, as the effective inertia of the grid decreases making the frequency more susceptible to perturbations \cite{milanoFoundationsChallengesLowInertia2018}.

The realistic modelling of frequency dynamics in large-scale power systems is profitable but complex due to its non-autonomous character. Stochastic dynamical models have successfully reproduced central characteristics of frequency measurements such as their non-standard distributions \cite{vorobevDeadbandsDroopInertia2019, gorjaoDatadrivenModelPowergrid2019, kraljic2022towards}. Such models can be used to generate synthetic frequency time series, which are, for example, employed to optimise electric devices \cite{guoSpectralModelGrid2021}. Moreover, they can be used to explore dynamics under different operating conditions, e.g., with an increased wind power generation \cite{ColetDynamicalModelFrequency}. However, multiple technical, economic and social influences and perturbations shape power system dynamics, as explained above. As a consequence, the power system must be regarded as a stochastic non-autonomous dynamical system, which makes grid frequency modelling a daunting task. 

In this context, the data-driven representation of external drivers can greatly facilitate realistic models, but data assimilation is challenging due to insufficient data sources. Integrating actual load time series can improve stochastic models of grid frequency dynamics in Continental Europe \cite{gorjaoDatadrivenModelPowergrid2019}. The assimilation of load and generation data enabled an accurate reproduction of grid frequency recordings for the Gran Canaria island \cite{ColetDynamicalModelFrequency}. However, load and generation time series are typically only available at hourly time scales \cite{ENTSOETransparencyPlatform}, while frequency dynamics happen at much smaller time scales, thus requiring a careful adoption of these external drivers. For large-scale power systems such as the Continental European grid, the data is often incomplete with missing or unrealistic data points \cite{hirthENTSOETransparencyPlatform2018}. The physical model is uncertain as for example control schemes vary among local control zones and detailed setups are not publicly available \cite{entso-eaisblReportDeterministicFrequency2019}. Finally, important dynamical parameters such as the inertia cannot be calculated exactly due to scarce time series on the power plant level \cite{entso-InertiaReport}. 

In this work, we propose physics-inspired machine learning (PIML) to approach these challenges. Compared to numerical simulations, PIML models can perform better in solving ill-posed problems with noisy insufficient data and imperfect physical models \cite{karniadakis2021physics, carleo2019machine}. Moreover, they can better generalise from small amounts of data than common machine learning methods, and they efficiently solve inverse problems of differential equations in situations with insufficient data or incomplete models \cite{karniadakis2021physics}. Notably, inverse problems are particularly important for power system control to estimate hidden states or dynamical parameters from measurements \cite{zhaoPowerSystemDynamic2019}. The inverse problem of inferring system parameters from input/output data is known as system identification \cite{ljungPerspectivesSystemIdentification2010} and PIML models offer a promising tool for such applications \cite{stiasnyPhysicsInformedNeuralNetworks2021}.

In particular, we develop a PIML model for the load-frequency dynamics of electric power systems, which includes proportional and integral controllers, stochastic noise and external techno-economic driving factors. The internal dynamics is described by a set of stochastic differential equations, which admit an analytic solution. The external driving is manifested through specific system parameters, which depend on a variety of techno-economic features such as the generation mix. This dependency is deduced via a feed-forward artificial neural network (FFNN), which is trained on data of the Continental European power system in a maximum likelihood approach. Finally, we interpret our model with SHapely Additive eXplanation (SHAP) values \cite{lundbergLocalExplanationsGlobal2020a, kruse2021revealing} to extract the dependency between dynamical parameters and techno-economic features. All in all, the model bridges the gap between the large-scale behaviour of interdependent energy systems and markets and the short-term dynamics of the power system. 

The article is organised as follows. In Sec.~\ref{sec:methods} we introduce the physics-inspired machine learning model for the grid frequency dynamics and discuss its implementation. In Sec.~\ref{sec:results} we present and evaluate three model applications: probabilistic prediction, system identification and explanation, and generation of synthetic time series. Finally, we discuss our results as well as possible future directions in Sec.~\ref{sec:discussion}.

\section{An integrated model for power system load-frequency dynamics}
\label{sec:methods}

Here, we present the details of how we constructed a physics-inspired model of the power grid frequency including a stochastic description of the frequency dynamics on coarse scales and the interaction with techno-economic features. Furthermore, we interpret the model in terms of power system operation and discuss the implementation as an artificial neural network. The detailed implementation of our data preparation and model pipeline, as well as all input data and the results are available on Zenodo \cite{piml_repo, kruseDataSetPIML}. 

\subsection{Short-term dynamics and control of the grid frequency}

Our starting point is a stochastic model for the dynamics of the grid frequency as illustrated in Fig.~\ref{fig:method}a. The rate of change of the frequency at a time $t$ is determined by the balance of power generation and load as well as the load-frequency control system (details are provided in appendix~\ref{app:aggregate_swing_eq}). Denoting the deviation from the reference as $\omega(t) = 2 \pi (f(t) -  f_{\rm ref})$, we have the equation of motion
\begin{equation}
    M \frac{\d \omega}{\d t} = P_{\rm im}(t) + P_{\rm noise}(t) + P_{\rm control}(t),
    \label{eq:equation-of-motion}
\end{equation}
where $M$ is the aggregated inertia constant. The power imbalance on the right-side has been decomposed into three contributions. The term $P_{\rm im}(t)$ denotes sustained power imbalances, for instance due to a mismatch of the load and the scheduled generation of dispatchable power pants (cf.~Fig.~\ref{fig:method}b). For time intervals of a quarter-hour, we can approximate the time dependence of these imbalances by an affine linear function 
\begin{align}
    P_{\rm im}(t) = M \cdot (q + rt),
    \label{eq:power-imabalance}
\end{align}
where $q$ models the power step of scheduled generation and $r$ represents the continuous drift of the load (cf.~Fig.~\ref{fig:method}c). The term $P_{\rm noise}(t)$ describes short-term fluctuations of the power balance, which we modelled as
\begin{align}
    P_{\rm noise}(t) = M \cdot D \cdot \xi(t), \label{eq:power-noise}
\end{align}
where $\xi(t)$ is white noise with a standard normal distribution and $D$ quantifies the strength of short-term power fluctuations. The term $P_{\rm control}(t)$ denotes the balance of primary and secondary load-frequency control, which can be modelled by a proportional-integral law as
\begin{align}
    P_{\rm control}(t) = - \frac{M}{\tau} \omega(t) 
      - \frac{M}{\kappa^2} \underbrace{\int^t_{t_i} \omega(t') \, \d t'}_{=: \theta(t)} 
      \label{eq:control-law}
\end{align}
with time constants $\tau$ and $\kappa$. We note that some simplifications are necessary to keep the model tractable. For instance, Eq.~\eqref{eq:control-law} neglects the existence of a small deadband in the proportional control law.

Due to the presence of noise, the equation of motion \eqref{eq:equation-of-motion} must be interpreted as a stochastic differential equation (SDE) with the explicit form
\begin{align}
    \ddt{\theta} &= \omega   \nonumber\\
    \ddt{\omega} &= q + rt - \omega \tau^{-1} -  \theta \kappa^{-2}  + D \xi(t). \label{eq:sde}
\end{align}
The parameter $\tau$ quantifies the effective primary control time scale. Equation \eqref{eq:sde} resembles a driven harmonic oscillator with an eigenfrequency $\kappa$, which can be interpreted as the intrinsic time scale of secondary control. In contrast, the effective time scale of secondary control is approximated by $\tau/\kappa^2$ \cite{vorobevDeadbandsDroopInertia2019}, as the frequency decays with this time constant in the overdamped case \cite{gorjaoDatadrivenModelPowergrid2019}. Note that these are only effective parameters which were rescaled by the inertia $M$. For example, the actual primary control strength is $M/\tau$ (Eq.~\eqref{eq:control-law}). However, the whole model is invariant under a scaling of $M$ (cf.~appendix~\ref{app:sde}) such that it is only possible to estimate the ratio of parameters and the inertia.

Applying It\^o's calculus, the SDE can be recast into a Fokker-Planck equation (FPE) of the probability density function $\PP(\omega,\theta;t)$
\begin{align}
    \frac{\partial}{\partial t} \mathcal{P}(\theta, \omega; t) 
    &= \bigg[ -\frac{\partial}{\partial \omega}
    \left(  q+rt - \tau^{-1} \omega - \kappa^{-2} \theta    \right)
    \nonumber \\
    & \qquad \quad  - \frac{\partial}{\partial \theta} \omega
    + \frac{D^2}{2} \frac{\partial^2}{\partial \omega^2}
    \bigg]
    \mathcal{P}(\theta, \omega; t)    .
    \label{eq:main-fpe}
\end{align}
As we show in appendix~\ref{app:FPE-sol}, the FPE is solved by a multivariate Gaussian distribution
\begin{align}
    \mathcal{P}(\v x; t) 
    = \frac{1}{2\pi | \v \Sigma |} 
    \exp\left(  -\frac{1}{2} (\v x -\v \mu)^\top 
    \Sigma^{-1} (\v x -\v \mu) \right)
    \label{eq:pdf-gauss}
\end{align}
with $\v x^\top = (\theta,\omega)$ and time-dependent parameters
\begin{align*}
    \v \mu(t) = \begin{pmatrix}
    \mu_\theta(t) \\ \mu_\omega(t)
    \end{pmatrix},
    \quad
    \v \Sigma(t) = \begin{pmatrix}
    \sigma^2_\theta(t) & \sigma_{\theta \omega}(t) \\
    \sigma_{\theta \omega}(t) & \sigma^2_\omega(t) 
    \end{pmatrix} \, ,
\end{align*}
if the parameters satisfy the ordinary differential equations
\begin{align}
    \frac{d}{dt} \mu_\theta &= \mu_\omega \nonumber \\
    \frac{d}{dt} \mu_\omega &= q+rt - \tau^{-1} \mu_\omega - \kappa^{-2} \mu_\theta \nonumber \\
    \frac{d}{dt} \sigma^2_\theta &= 2 \sigma_{\theta \omega} \nonumber \\
    \frac{d}{dt} \sigma^2_\omega &= \sigma^2\omega - \tau^{-1}  \sigma_{\theta \omega} - \kappa^{-2} \sigma^2_\theta \nonumber \\
    \frac{d}{dt} \sigma_{\theta \omega} &= -2 \tau^{-1} \sigma^2_\omega -2 \kappa^{-2} \sigma_{\theta \omega} \, .
    \label{eq:eom-musigma}
\end{align}
Here, $\mu$ and $\sigma$ are the mean and standard deviation of the angle and the frequency deviation, while $\sigma_{\theta \omega}$ represents their covariance. 

In appendix~\ref{app:sol_moment_eq}, we solve the moment equations \eqref{eq:eom-musigma} analytically, thus solving the entire stochastic dynamic model, which vastly simplifies the analysis.

\subsection{Power system operation and interdependecies}

\begin{figure*}[tb]
    \centering
    \includegraphics[width=\textwidth]{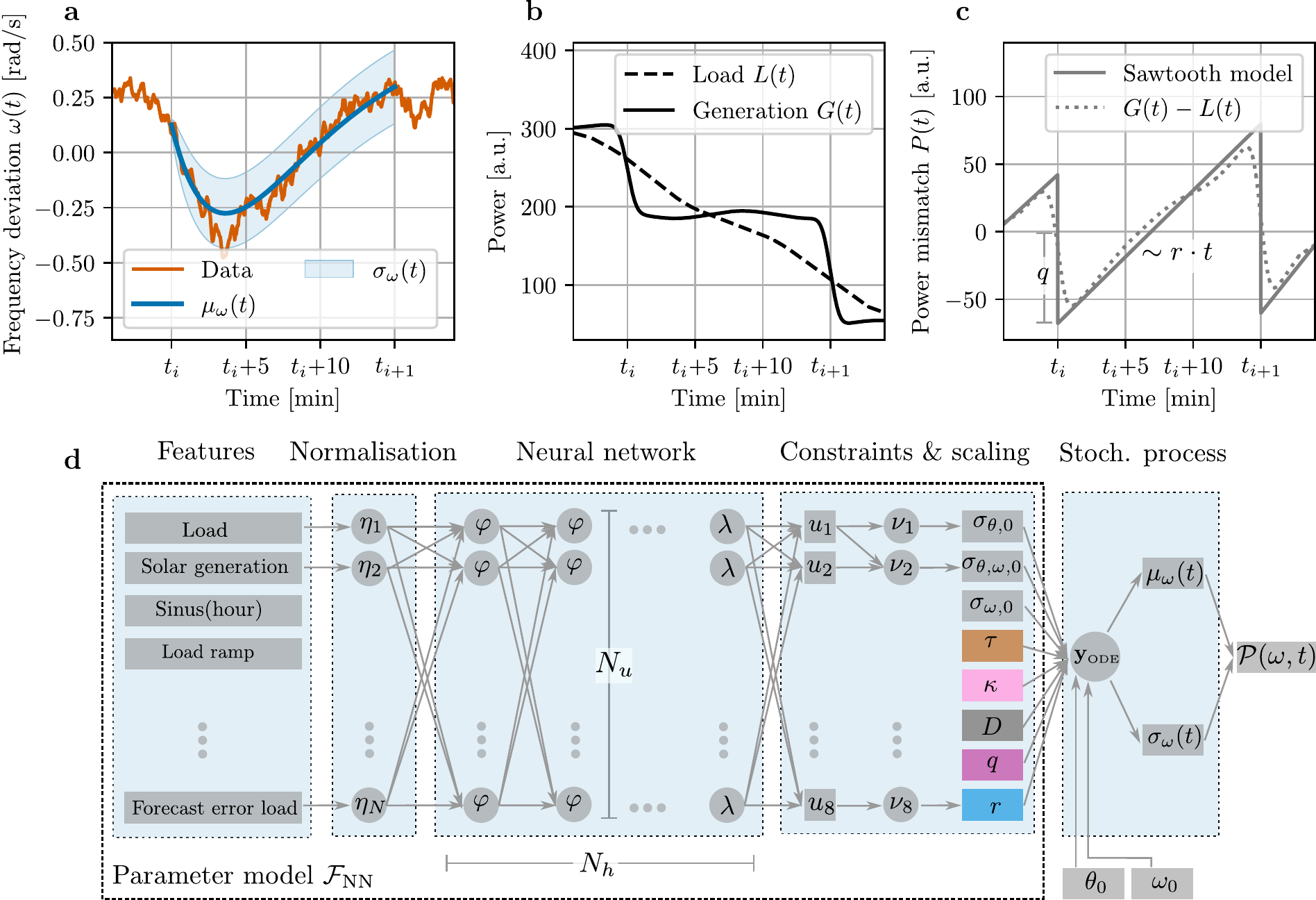}
    \caption{
    A physics-inspired machine learning model for the power grid frequency.
    \textbf{a} In each interval $[t_i, t_{i+1}]$, we modelled the grid frequency with a stochastic process that provides a normal distribution with time-dependent mean $\mu_{\omega}(t)$ and standard deviation $\sigma_{\omega}(t)$. The panel depicts our model prediction of the Continental European grid frequency at $t_i=$ 23:00 on the 28th of October 2018 in comparison to the recorded data.   
    \textbf{b,} Deviations from the reference frequency are partly driven by deterministic power imbalances $P_{\rm im}(t)=M\cdot P(t)$. These result from a different time evolution of generation $G(t)$ and load $L(t)$ due to the market-based dispatch of power generation, which is illustrated in this panel using self-engineered synthetic data. \textbf{c,} We approximate the deterministic mismatch with a sawtooth function $P(t)$.  \textbf{d,} The full probabilistic model $\mathcal P(\omega,t)$ incorporates deterministic imbalances $P_{\rm im}(t)$, additional stochastic imbalance fluctuations $P_{\rm noise}(t)$ and the load-frequency control $P_{\rm control}(t)$. The parameter model $\mathcal{F}_{\rm NN}$ predicts the model parameters, i.e., the imbalance and control parameters (colour highlight), as well as the initial covariances, from $N=77$ techno-economic features by using a feed-forward neural network (FFNN). 
    }
    \label{fig:method}
\end{figure*}

The SDE for the frequency dynamics in Eq.~\eqref{eq:sde} contains several parameters, describing the load-frequency control system ($\tau$, $\kappa$), or the power imbalances on different time scales ($r$, $q$, $D$). The parameters are not constant, but change during the day. For instance, the market based scheduling of conventional power plants causes characteristic imbalances of generation and load \cite{weissbachHighFrequencyDeviations2009,kruse2021exploring}. Electricity is traded on the spot markets in blocks of 15, 30 or 60 minutes, leading to characteristic patterns of the power imbalance $P_{\rm im}(t)$ illustrated in Fig.~\ref{fig:method}b,c. The shape of these patterns, as well as other properties of the power system, change in time due to the influence of a variety of techno-economic features.

We thus propose a model that integrates the internal dynamics of the frequency-control system, the stochastic noise, and the impact of various techno-economic features. In every 15 minute interval, the frequency is modelled by the SDE \eqref{eq:sde}. The system parameters $\tau$, $\kappa$, $D$, $r$, $q$ change from interval to interval, depending on the influence of external techno-economic features as detailed below. This dependence is modelled by a FFNN, that is trained such that the stochastic dynamics best fits the recorded time series. More precisely, the FFNN constitutes a parameter model $\mathcal F_{NN}: \boldsymbol{X} \mapsto \boldsymbol{\vartheta}$, where $\boldsymbol{X}$ summarises the values of techno-economic features (Fig.~\ref{fig:method}d). The vector $\boldsymbol{\vartheta}$ includes the system parameters $\tau$, $\kappa$, $D$, $r$, $q$ as well as the initial covariances $\sigma_{\omega \theta, 0}$, $\sigma^2_{\omega,0}$ and $\sigma^2_{\theta,0}$ at time $t=t_i$, while the initial means $\mu_{\omega,0}$ and $\mu_{\theta,0}$ are directly obtained from the data.

We thus establish a model that links different temporal and technological scales, from the slow evolution of electricity markets to the fast dynamics of the power grid frequency. The integrated model predicts a probability distribution $\PP(\omega,\theta;t)$ for an entire interval of 15 minutes from techno-economic features.

\subsection{Techno-economic features affecting power system dynamics and operation}

As input features for the PIML model, we used several operational time series of the Continental European power system from the ENTSO-E Transparency platform \cite{ENTSOETransparencyPlatform}. Following ref.~\cite{kruse2021revealing}, we downloaded load forecasts, day-ahead scheduled generation, day-ahead forecast data for wind and solar power, day-ahead electricity prices, actual generation per type and actual load. In addition, we included pumped hydro consumption (cf. \cite{kruseSecondaryControlActivation2022a}), as well as net scheduled, i.e., market-based flows and net physical flows between Continental Europe and other synchronous areas (cf. \cite{putzRevealingInteractionsHVDC2022}).

We prepared the features for each 15 minute interval. This opens the possibility to capture the effects of intraday electricity markets that operate in 15 minute intervals, while the day-ahead electricity market typically acts every 60 minutes \cite{linElectricityMarketsTheories2017}. To this end, we determined the time resolution of each feature in each country, which can vary due to different market designs. Then, we upsampled 60 minute data using linear interpolation for load and renewable generation data, and forward padding for all other feature types. 

To gain additional interpretable input data, we engineered physically meaningful features and aggregated the data area-wise (cf. \cite{kruse2021revealing,kruseSecondaryControlActivation2022a}). We included forecast errors (day-ahead minus actual), ramps (time derivative of a feature) and unscheduled flows (scheduled minus physical flows). Since all features are only available on country-level, the data was finally aggregated within the whole Continental European area to represent the aggregated impact on the grid frequency (the detailed implementation is available on Zenodo \cite{piml_repo}). 

The grid frequency recordings used in this work were taken from ref.~\cite{krusePreProcessedPowerGrid2020a}, which provides pre-processed frequency data from the German transmission system operator TransnetBW \cite{transnetbwfreqdata}.

\subsection{Artificial neural network model}

The integrated model provides a probabilistic prediction of the power grid frequency for every 15-minute interval $[t_i, t_{i+1}]$. The architecture of the model is depicted in Fig.~\ref{fig:method}d.  

As input, we used the techno-economic features $\v X(i) = (X_1(i),..., X_N(i))^T$ for each time interval $[t_i, t_{i+1}]$. In the first step, each feature $X_{k}$ was normalised using functions $\eta_k(X_{k}) = (X_{k} - \langle X_{k} \rangle) / \sigma_{{k}}$ to improve numeric stability, where $\langle \cdot \rangle$ denotes the average and $\sigma_k$ the standard deviation of the feature. 

The normalised features were fed into a FFNN of $N_h$ hidden layers with $N_u$ units and activation functions $\varphi$. The last layer comprises a linear activation $\lambda$, as we aim to predict real-valued parameters $\v \vartheta$. 

The following layer rescales the output of the FFNN and implements several constraints. The rescaling was implemented to improve training efficiency and stability. After random initialisation, the outputs $u_j$ of the FFNN typically have the same scale, but the physical parameters do not. Such a mismatch will yield large initial errors along certain parameter axis leading to inhomogeneous loss landscapes which can make optimisation inefficient and more difficult \cite{mehtaHighbiasLowvarianceIntroduction2019}. This difficulty can be mitigated by a suitable rescaling.
Furthermore, several output variables must respect physical constraints. For instance, the time constants $\tau$ and $\kappa$ must be positive and respect the inequality $\kappa \ge 2\tau$ to avoid an unphysical oscillation behaviour of the solution. Rescaling and constraints were implemented with parameter-specific functions $\nu_j(u_j)$, which are described in appendix~\ref{app:scalings}.

After rescaling, the output $\v \vartheta(i)$ of the parameter model $\mathcal{F}_{\rm NN}$ was used to compute a probabilistic prediction of the grid frequency for the entire time interval $[t_i,t_{i+1}]$ based on Eq.~\eqref{eq:pdf-gauss}. The vector $\v \vartheta(i)$ contains the system parameters as well as the covariances at $t=t_i$, while the means $\mu_{\omega,0}$ and $\mu_{\theta,0}$ are directly taken from data. For training and forecasting applications, we used the actual value of the frequency $\mu_{\omega,0}(i) = \omega(t_i)$ and estimated $\mu_{\theta,0}(i)= \int^{t_i}_{t_i - 60s} \omega(t') \, \d t'$. For the generation of synthetic time series, we predicted intervals sequentially in time and estimated $\mu_{\omega,0}$ and $\mu_{\theta,0}$ from the preceding prediction and not from the data.

\subsection{Training, testing and interpretation}
Our complete data set comprises 107650 data points from 2015 to 2019. In particular, it includes features $\v X(i)$ and frequency time series $\v \omega(i) = (\omega(t_i), ..., \omega(t_i+t_{\rm max}))^T$ for each interval $i$. To assess the time-dependence of the performance, we modelled and predicted subsets of the 15 minute interval with $t_{\rm max}<$ 15 min (cf.~Sec.~\ref{sec:performance}), but the full interval ($t_{\rm max}=$ 15 min) was used in all other cases.

We quantified the ability of the model to predict the stochastic frequency dynamics by the negative log-likelihood. For a given time interval $\mathcal I = [t_i, t_i+t_{\rm max}]$, the negative log-likelihood is defined as 
\begin{align}
    \mathcal C(\v \omega(i), \v \vartheta(i)) = 
    -\sum_{t \in \mathcal I}
    \log \mathcal P(\omega; t | \v \vartheta(i) ), 
    \label{eq:loglike}
\end{align}
where $P(\omega; t | \v \vartheta )$ is the marginal of the PDF \eqref{eq:pdf-gauss} evaluated at the measured data $\omega(t)$. The log-likelihood is a negatively oriented metric, i.e., smaller values represent a better performance.

To train the FFNN, we initialised the weights using the Glorot uniform initialiser \cite{glorotUnderstandingDifficultyTraining2010}. Using data from 2015 to 2017, we trained the weights with stochastic gradient descent using the ADAM optimiser with a fixed learning rate \cite{mehtaHighbiasLowvarianceIntroduction2019}. As a loss function we chose the negative log-likelihood \eqref{eq:loglike}, summed over all quarter-hour intervals in the training set.
The model hyperparameters were optimised using random search on data from 2018 (as a validation set) and with parameter choices defined in Tab.~\ref{tab:hps}. In particular, we trained the model for 100 epochs and applied early stopping based on the validation loss. Then, we retrained the best model on data from 2015 to 2018 and evaluated the performance in terms of the negative log-likelihood on data from 2019 as a test set.

\begin{table}[tb]
    \caption{Parameter choices during hyperparameter optimisation. $\textrm{Sig}(x)$ is the sigmoid function and $\tanh(x)$ the hyperbolic tangent.     } 
    \label{tab:hps}
    \centering
    \begin{tabular}{c|c}
         & Possible values  \\ \hline
         Learning rate & $10^{-4}$, $10^{-3}$, $10^{-2}$ \\ \hline
         Dropout rate & 0, 0.1, 0.2, 0.3 \\ \hline
         $N_u$ & 64, 128 \\ \hline
         $N_h$ & 3,5,7\\ \hline
         Activation $\varphi(x)$ & $\textrm{Sig}(x)$, $\tanh(x)$
    \end{tabular}
\end{table}

We benchmarked the developed model by comparing its performance to the daily profile of the grid frequency, which is defined as follows. For a fixed time of the day $t_d$, we collected all frequency values recorded on all days in the training set and calculated their average $\mu_p(t_d)$ and the corresponding standard deviation $\sigma_p(t_d)$. Our daily profile model $\mathcal P_p$ returns a normal distribution $\mathcal P_p(\omega;t) = \mathcal N(\mu_p(t_d), \sigma_p(t_d))$ based on the time of the day $t_d(t)$ of the time step $t$. For example, the predicted mean $\mu_p(t_d)$ for January 11, 2019, at 11:00 equals the average of frequency values at 11:00 over all days in the training set.  In addition to the daily profile, we applied the constant model as a benchmark, which simply provides a normal distribution using the global mean and variance of the whole frequency time series.  

Finally, we interpreted our parameter model $\mathcal F_{NN}: \boldsymbol{X} \mapsto \boldsymbol{\vartheta}$ with SHapely Additive eXplanation (SHAP) values \cite{lundbergLocalExplanationsGlobal2020a}, which attribute the prediction of a single parameter $\vartheta_{j}(i)$ to the impact of different features $X_k(i)$. Aggregating individual SHAP values offers a tool to inspect feature importances and dependencies extracted by the FFNN. In particular, we used KernelSHAP \cite{lundbergUnifiedApproachInterpreting2017}, which approximates SHAP values for any machine learning model. 

Our FFNN model is implemented with tensorflow \cite{abadi2016tensorflow} and tensorflow probability \cite{dillonTensorFlowDistributions2017} and we used keras tuner for hyperparameter optimisation \cite{omalley2019kerastuner}.

\section{Model application and evaluation}
\label{sec:results}

We demonstrate and evaluate three applications of our PIML model. First, it provides a probabilistic prediction of the grid frequency trajectory in each time interval, which we evaluate in terms of the performance and compare it to elementary benchmarks (Sec.~\ref{sec:performance}). Second, the model infers time-dependent imbalance and control parameters based on the data, i.e., we can use it for system identification. We analyse their time-dependence, compare our estimates with values from the literature and explain their dependency on techno-economic features with SHAP values (Sec.~\ref{sec:inference}). Third, our model provides a tool for generating synthetic frequency time series by drawing samples from the stochastic process. Such synthetic scenarios should reproduce central stochastic characteristics of the grid frequency, which we evaluate in Sec.~\ref{sec:scenario}.

\subsection{Probabilistic prediction of the grid frequency}
\label{sec:performance}

\begin{figure*}[tb]
    \centering
    \includegraphics[width=\textwidth]{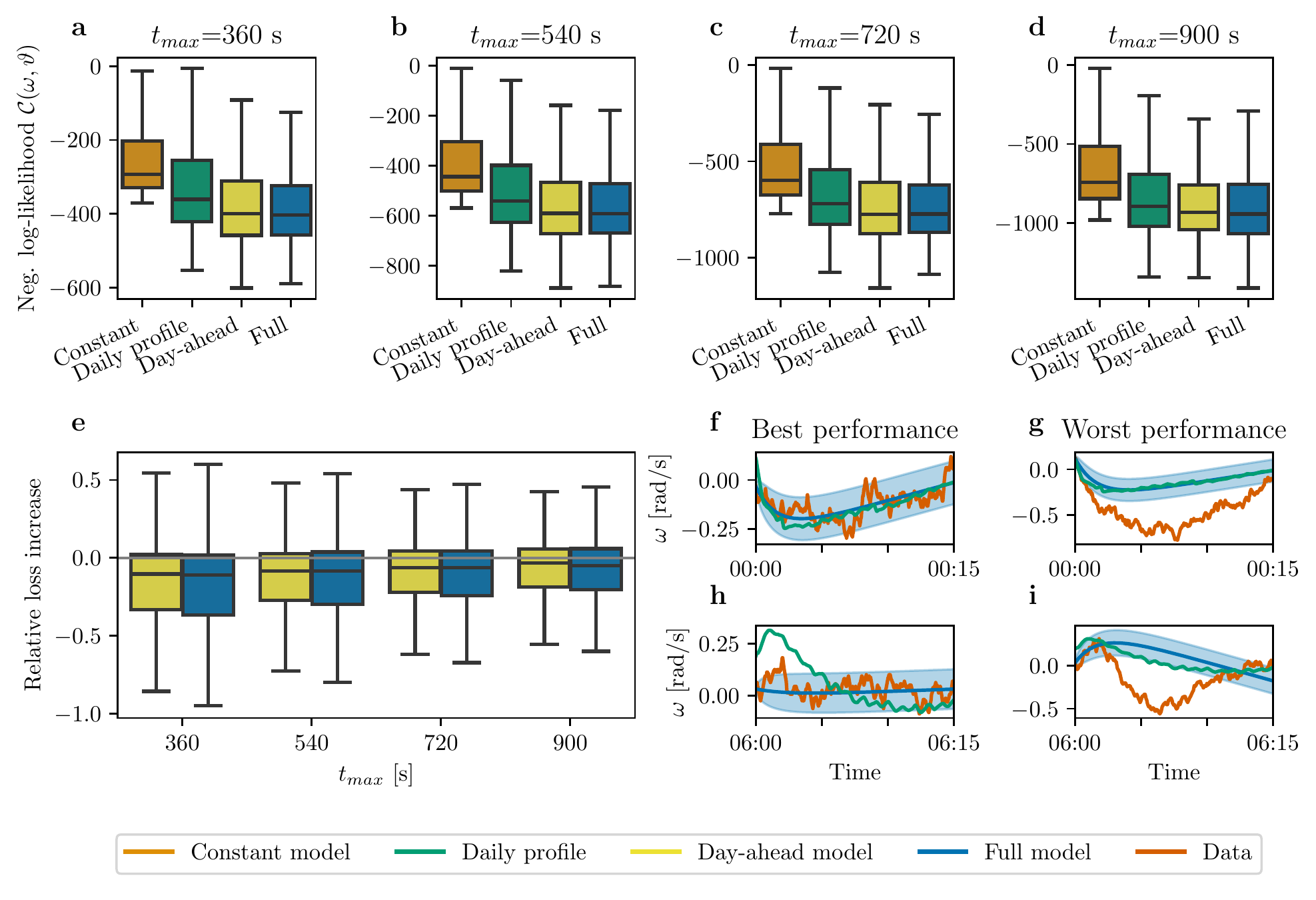}
    \caption{Probabilistic machine learning model outperforms elementary benchmarks. \textbf{a,b,c,d,} We quantify the performance of our probabilistic frequency prediction with the negative log-likelihood loss, which implicates better performance if values are lower. Irrespective of the prediction length $t_{\rm max}$, the day-ahead and full models outperform the the benchmark models, namely the constant model and the daily profile predictor. \textbf{e,} The relative loss increase between our model and the daily profile provides a comparison between different prediction lengths, yielding a slight decrease of the performance with longer predictions. \textbf{f,g,h,i,} Prediction examples with best and worst performance at 00:00 and 06:00 illustrate the strengths and limitations of our model.}
    \label{fig:performance}
\end{figure*}

Our physics-inspired model provides a probabilistic prediction for unseen samples of the grid frequency. Its performance depends both on the length of the prediction and on the available set of features (Fig.~\ref{fig:performance}). 

The machine-learning model outperforms elementary benchmarks irrespective of the prediction horizon (Fig.~\ref{fig:performance}a-d). 
Using all available techno-economic features, the full model yielded lower median loss values than the daily profile and the constant model for each prediction length $t_{\rm max}$. Restricting the feature set to day-ahead available data yields a similar performance, which enables us to forecast future frequency deviations better than the daily profile.

However, the performance slightly deteriorates with increasing prediction length (Fig.~\ref{fig:performance}e). The absolute log-likelihood of different data sets, i.e., of different $t_{\rm max}$ values, cannot be compared. We therefore employed the relative loss increase between the machine learning model and the daily profile as a measure, which mainly exhibits negative values as our model outperforms the daily profile. Predicting only the first $t_{\rm max}=360$ seconds yielded a better performance then predicting the full interval ($t_{\rm max}=900$ s). This points to a potential limitation of our model at the end of the prediction interval, which is likely due to the approximate treatment of the power imbalance $P_{\rm im}(t)$ (cf.~Fig.~\ref{fig:method}c). Our model assumes a discrete step at the start of an interval, while real power plants start ramping up or down continuously earlier at the end of the previous interval  \cite{entso-eaisblReportDeterministicFrequency2019}. Hence, the frequency $\omega(t)$ at the end of an interval is already driven by the dispatch in the future interval. This aspect is not included in the physics-inspired model, while it is present in the daily profile.  

The prediction examples in Fig.~\ref{fig:performance}f-i illustrate the strengths and limitations of our model. The intervals with the best model performance at 00:00 and 06:00 demonstrate how our model outperforms the daily profile by far. A remarkable aspect is observed when inspecting the intervals with the worst performance: As the physics-inspired model fails to capture the dynamics, so does the daily profile, albeit at a different magnitude (Fig.~\ref{fig:performance}g,i). The limitations of our model due to continuous generation ramps turned up in Fig.~\ref{fig:performance}i: The frequency increased and then ramped down expectedly due to the rising load in the morning, which causes upwards deterministic frequency deviations. However, the frequency ramped up again towards the end of the interval, which is not covered in our model (see above), but slightly visible in the daily profile. 

In the following sections, we explain dynamical parameters based on techno-economic features, among others. The full model better suits for explanation as it also includes actually measured features, such as forecast errors. Therefore, we only focus on the full model and predict the full interval (with $t_{\rm max} = 15$ min) in following sections.

\subsection{System identification and explanation}
\label{sec:inference}

\begin{figure*}[tb]
    \centering
    \includegraphics[width=\textwidth]{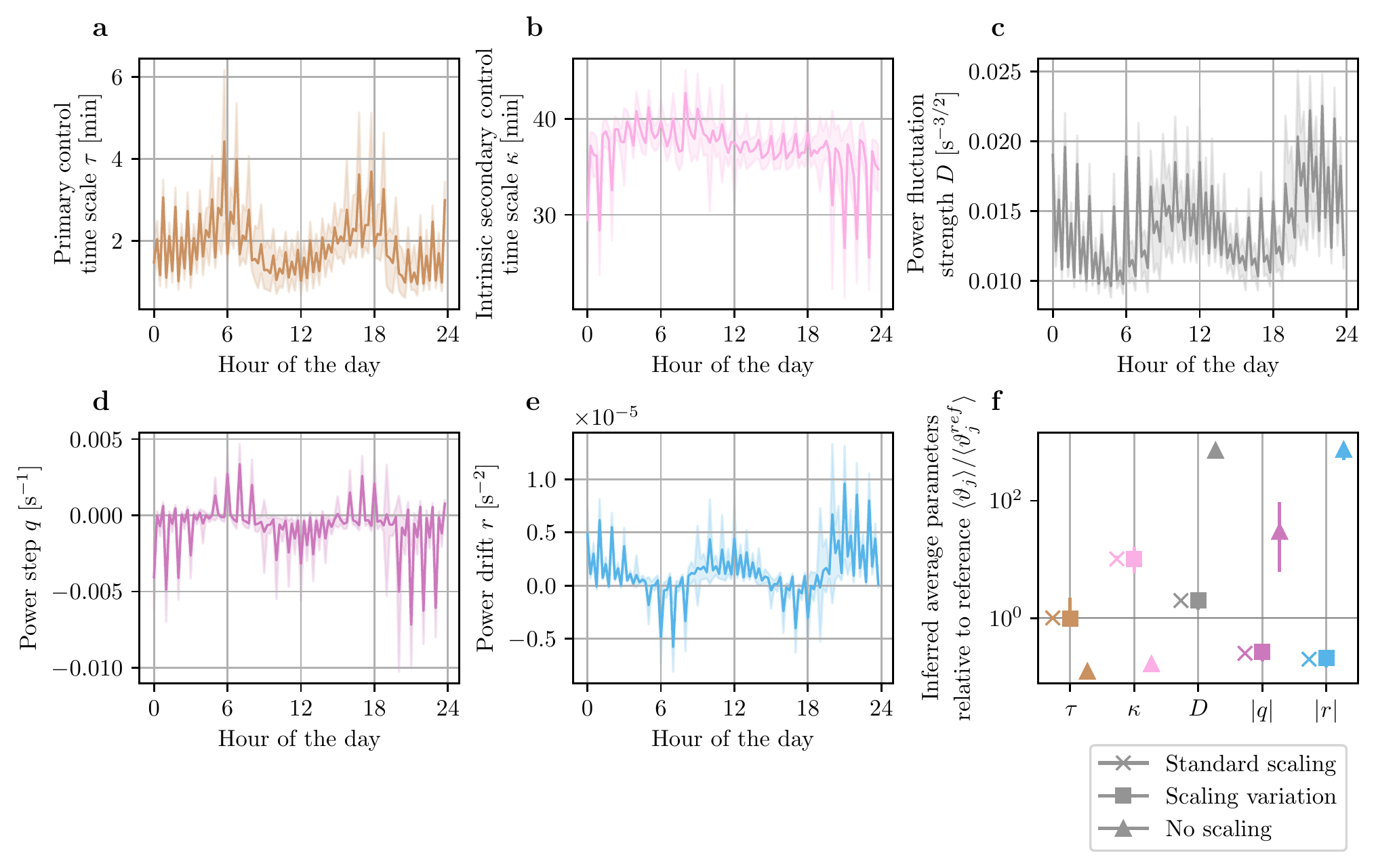}
    \caption{Physics-inspired machine learning model infers dynamical system parameters. \textbf{a,b,c,d,e} The inferred parameters show strong daily patterns, which underlines the importance of time-dependent dynamical properties in our model. The panels depict the daily means (solid lines) and the range between the 25~\% and 75~\% quantile (area). \textbf{f,} The ratio between the average inferred parameters and the reference value $\vartheta_j^{\rm ref}$ obtained from time-independent stochastic inference (cf.~appendix~\ref{app:param-ref-values}). In addition to the standard scaling used in the main model and a model with no scaling, we tested small variations of the scaling parameters (defined in Tab.~\ref{tab:constraints}), with 10 random weight initialisations for each combination of scaling parameters. The ensemble means (markers) and the data range (errorbar) indicate the relation between the inferred parameters and the reference, which shows particularly large deviations if no scaling is applied. 
    } 
    \label{fig:parameters}
\end{figure*}

\subsubsection{Inference and variation of system parameters}

In addition to probabilistic prediction, our PIML model provides a tool to infer dynamical system parameters from frequency measurements and techno-economic features (Fig.~\ref{fig:parameters}). In contrast to time-independent models \cite{anvariStochasticPropertiesFrequency2019, gorjaoDatadrivenModelPowergrid2019}, our parameter model $\mathcal F_{NN}$ extracts time-dependent system parameters $\v \vartheta(i)$ for each time interval $i$, which mirror the local dynamical properties of load-frequency control (cf.~Fig.~\ref{fig:method}). Note that we only estimate effective parameters that also contain the impact of the inertia (cf.~Sec.~\ref{sec:methods}), which we discuss later in Sec.~\ref{sec:discussion}.

The inferred parameters strongly change during the day, which illustrates the importance of time-dependent dynamical modelling (Fig.~\ref{fig:parameters}a-e). The daily profile of the primary control time scale shows variations of 14~\%, while the intrinsic secondary control time scale varies by 16~\%. The power imbalance parameters show even stronger variations, with the short-term fluctuation strength $D$ varying by 61~\% and the deterministic parameters $q$ and $r$ changing by 550~\% and 442~\%.

The imbalance parameters show distinct patterns that reveal physically meaningful impact factors on the grid frequency. We inferred upwards power steps $q$ and negative drifts $r$ in the morning around 06:00 and in the evening around 18:00, while the opposite behaviour was estimated around noon and during the night. This successfully models the deterministic imbalances between scheduled generation and continuous load: Around noon and during the night, the load is decreasing thus causing downward power steps and positive drifts (cf.~Fig.~\ref{fig:method}b,c). Moreover, the inferred power steps $q$ peak at the beginning of the hour, while being smaller during the rest of the hour. This indicates the dominance of generation changes within 1 h intervals over changes in 15 min periods, which is consistent with the characteristics of the European electricity markets: In fact, much more generation volume is traded on the hourly day-ahead market then on the quarter-hourly intraday market in Europe \cite{IntradayTradingIncreases}.  

The inferred parameters generally agree with estimates from the literature \cite{gorjaoDatadrivenModelPowergrid2019}, with small differences due to redundancies between secondary control and deterministic deviations (Fig.~\ref{fig:parameters}f). The panel depicts the ratio between the time average of our absolute parameter estimates and the reference value from the literature $\vartheta_j^{\rm ref}$ (cf. appendix~\ref{app:param-ref-values}), which was extracted by a time-independent model. Our model (with standard scaling defined in appendix \ref{app:scalings}), inferred values for $\tau$ and $D$ that are very similar to the reference, i.e., the ratio to the reference is near one. The secondary control was weaker in our model (larger time scale) and the deterministic power mismatch was also weaker, i.e., our model inferred smaller absolute generation steps $q$ and drifts $r$. As discussed in ref.~\cite{gorjaoDatadrivenModelPowergrid2019}, there is a redundancy between secondary control and deterministic power drifts, because both can drive the frequency back to its reference. This makes it generally difficult to obtain unique estimates, which might explain the difference to the literature values for $\kappa$.

The adequate inference of dynamical parameters is greatly facilitated by our implementation of appropriate scaling steps. Fig.~\ref{fig:parameters}f depicts the parameter estimates if no scaling is applied (triangle markers), which resulted in much stronger deviations from the literature values. This is probably due to the large difference in scale between the parameters, which renders the FFNN training inefficient and unstable due to the very heterogeneous loss landscape \cite{mehtaHighbiasLowvarianceIntroduction2019}. Note that small variations of the scaling coefficients (defined in appendix~\ref{app:scalings}) did not strongly change the parameter estimates (square markers) such that the results seem to be independent of the exact choice of the scaling. 

\subsubsection{Techno-economic drivers of dynamical system properties}

\begin{figure*}[tb]
    \centering
    \includegraphics[width=\textwidth]{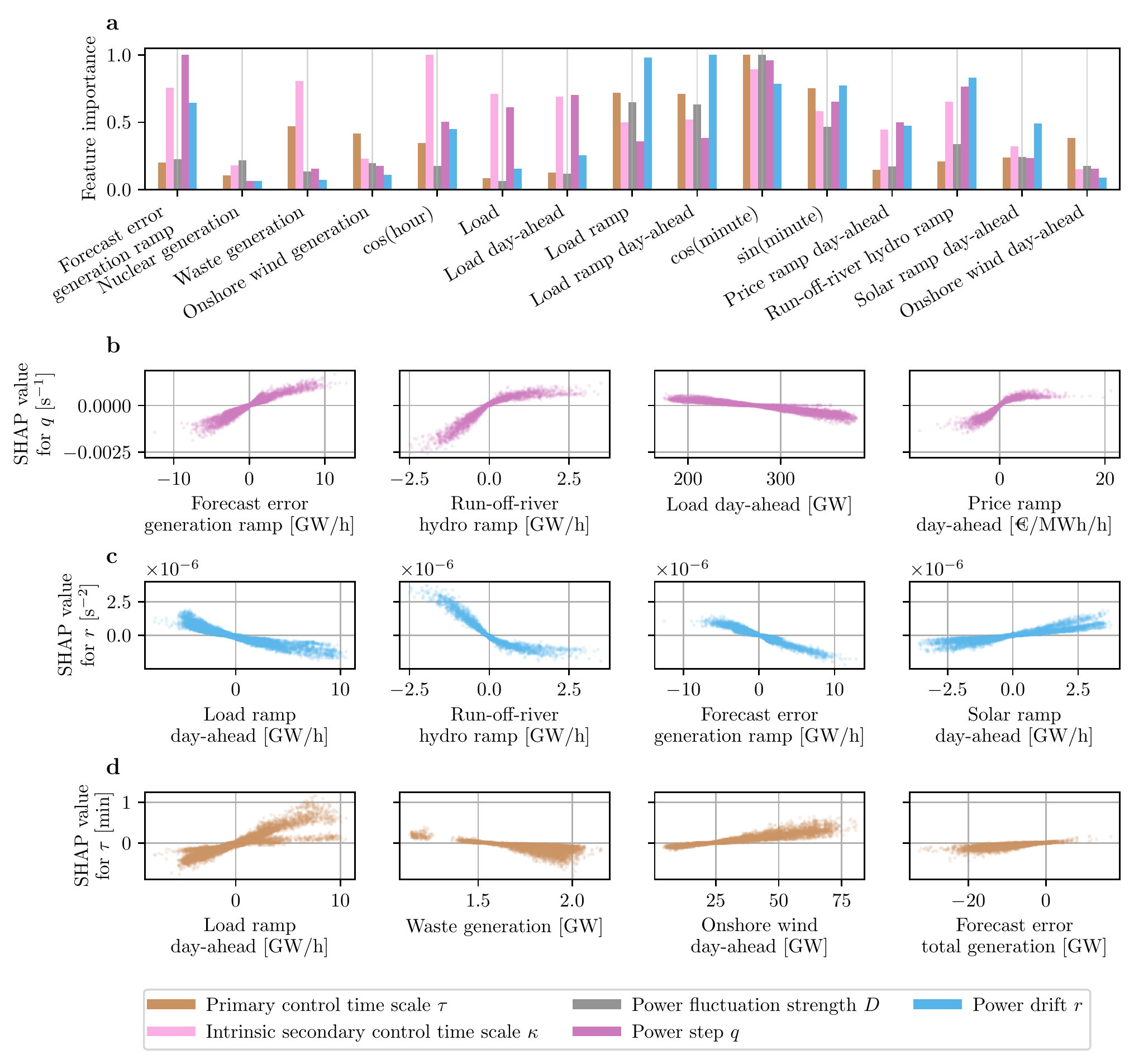}
    \caption{SHAP values reveal dependencies between techno-economic features and dynamical system parameters. \textbf{a,} The mean absolute SHAP values quantify the overall importance of a feature, which varies strongly among the different dynamical system parameters. \textbf{b,c,d,} The relation between feature values and SHAP values reveals the dependencies that were extracted by our model. The inferred dependencies generally represent physically meaningful effects that agree with domain knowledge (see main text). In this panel, we only display the four most important techno-economic features, i.e., we excluded the time encoding features such as sinus of the minute "sin(minute)". Before the selection, we also excluded the actual features, e.g., actual load ramps, if a day-ahead version exists, e.g., load-ramp day-ahead, because the dependencies of these actual and day-ahead available features were largely the same.}
    \label{fig:shap}
\end{figure*}

Using SHAP values \cite{lundbergLocalExplanationsGlobal2020a}, we explain the dependencies between the techno-economic features and the dynamical parameters $\vartheta_j$, which the model extracted from the data (Fig.~\ref{fig:shap}). Focusing on the deterministic mismatch parameters $q$ and $r$ and the primary control time scale $\tau$, we analysed feature importances quantified by the mean absolute SHAP value (Fig.~\ref{fig:shap}a), as well as dependencies (Fig.~\ref{fig:shap}b-d) that display SHAP values for different feature values.

The power step $q$ was mostly determined by generation ramps and forecast errors (Fig.~\ref{fig:shap}a,b). The most important feature, forecast error generation ramp, represents the difference between day-ahead scheduled generation ramps (on a 1h basis) and actual values of the total generation ramps that also include intraday trading within 15 minute intervals. Therefore, they mirror the additional 15 minute ramps that are not included in the day-ahead generation ramps thus making the feature essential for the model to estimate the power steps $q$.
In addition, it is known that especially fast generation ramps drive the power step and thus the rate of change of frequency (RoCoF) at the beginning of the market intervals \cite{kruse2021revealing}. Accordingly, the PIML model yielded a high importance of hydro power ramps, which are among the fastest in the European power system. Interestingly, positive ramps exhibited a smaller effect then negative ramps. This probably relates to the limitation of our model in representing upwards deterministic deviations. Upwards deviations typically start before the start of the interval (cf.~Sec.~\ref{sec:scenario}), while downward ramps rather follow the approximation of a discrete power step at $t=t_i$. Therefore, the remaining upward power step at $t_i$ represents only a part of the total step. Hence, the model sees a step which is smaller than for downwards deviations and thus assumes a weaker effect of positive hydro ramps on the step $q$. 

The drift $r$ of the deterministic mismatch mirrors continuous changes of the load (Fig.~\ref{fig:method}b,c). Consistently, load ramps obtained the highest feature importance for $r$ with positive load ramps leading to negative slopes $r$ (Fig.~\ref{fig:shap}a,c). Solar ramps were also ranked highly, but their dependency showed the opposite behaviour. This mirrors the fact that in addition to the load, solar power also shapes the slow evolution of the deterministic mismatch $P_{\rm im}(t) \sim r\cdot t$ \cite{kruse2021exploring}: The load and aggregated solar power typically change slowly and continuously on a time scale of hours, with the load having a negative impact and the solar power having a positive impact on the power imbalance. This perfectly manifests in the opposite effects of load and solar ramps on the mismatch slope $r$, which were identified by our PIML model (Fig.~\ref{fig:shap}c). 

The effective time scale of primary control $\tau$ was determined by load ramps, waste power generation and wind power (Fig.~\ref{fig:shap}a,d). Most interestingly, increasing wind power generation led to a larger time scale of primary control. In an Ornstein-Uhlenbeck process, $\tau$ quantifies the time to revert back to the mean after a disturbance. A large wind power feed-in can cause large stochastic imbalances and thus effectively reduce the mean-reverting time $\tau$. This would be consistent with our SHAP results and with previous studies that showed an increased variability of short-term frequency dynamics with increasing wind power feed-in \cite{haehneFootprintAtmosphericTurbulence2018}. Notably, this dependency cannot be caused by the rescaling of dynamical parameters with the inertia. If the actual primary control strength $M/\tau$ was constant (cf.~Sec.~\ref{sec:methods}), increasing wind power, and thus decreasing inertia $M$, would correspond to decreasing values of $\tau$. However, we observed the opposite: the dependency  showed increasing values of $\tau$ (Fig.~\ref{fig:shap}d), thus pointing to other causes such as an increased variability. 

Finally, note that the hour and minute features were very important, in particular for $\tau$, $\kappa$ and $D$ (Fig.~\ref{fig:shap}a). This points to missing information in the feature set, so that the model relies on an average daily behaviour using the hour and minute features. 

\subsection{Generation of synthetic grid frequency time series}
\label{sec:scenario}

\begin{figure*}[tb]
    \centering
    \includegraphics[width=\textwidth]{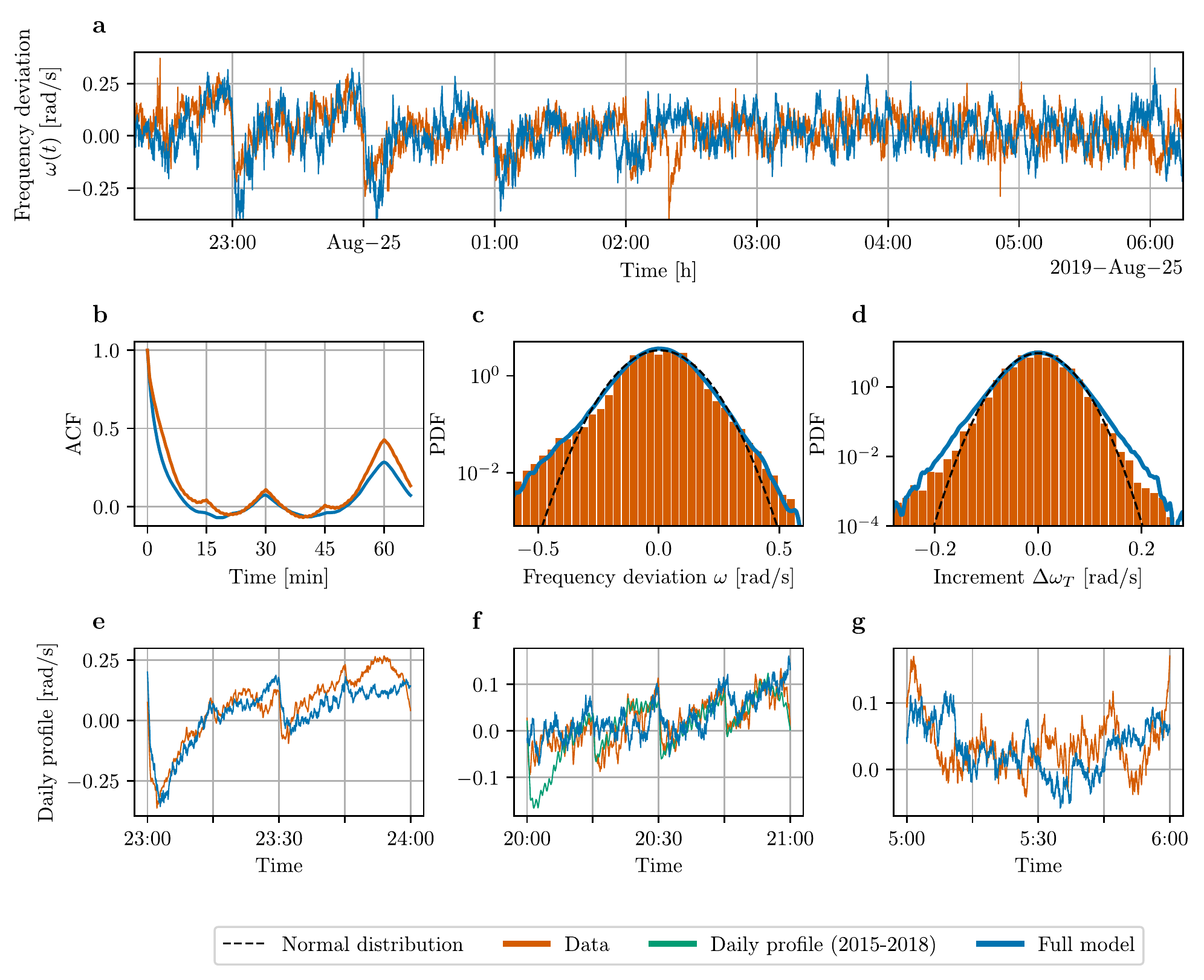}
    \caption{Physics-inspired machine learning model generates synthetic grid frequency time series with characteristic properties. \textbf{a,} We sampled a synthetic frequency trajectory from our probabilistic model for the test period August 25 to September 05 in our test set, which was the longest period without missing or corrupted data points. The panel depicts the start of this period showing a good agreement between real and synthetic time series. 
    \textbf{b,c,d,} The autocorrelation function (ACF) and the probability density function (PDF) of the frequency deviation $\omega(t)$ and its increments $\Delta \omega_T(t) = \omega(t+T) - \omega(t)$ ($T=10$ s) have heavy tails, which is well reproduced by our synthetic time series. 
    \textbf{e,f,g}. The daily average profiles of the real and synthetic frequency time series within the test period. For 20:00, we additionally display the daily profile computed from the training set (2015-2018). Our model better fits the frequency in this hour and therefore grasps local time-dependent properties of its daily evolution. 
    }
    \label{fig:scenarios}
\end{figure*}

A third major application of probabilistic machine learning models is the generation of synthetic time series. Scenario generation, i.e., the generation of multiple synthetic samples from the model, is important for simulation or optimisation models \cite{cramerValidationMethodsEnergy2022, cramer2022conditional}. Given a data set of external features, synthetic time series are obtained as follows. For every interval $i$, we applied the FFNN to predict the system parameters $\v \vartheta(i)$. We then integrated the original SDE \eqref{eq:sde} using a standard Euler-Maruyama method. To ensure continuity, we used the final values of $\omega$ and $\theta$ from one interval as initial states for the following interval. As a test case, we generated a synthetic trajectory from August 25 to September 05 in our test set, for which the first hours are shown in Fig.~\ref{fig:scenarios}a.

Power grid frequency trajectories exhibit several highly characteristic stochastic properties \cite{schaferNonGaussianPowerGrid2018}: The distribution of both frequency $\omega(t)$ and its increments $\Delta \omega_T(t) = \omega(t+T) - \omega(t)$ is heavy tailed (Fig.~\ref{fig:scenarios}b,c). Large deviations and large jumps are much more likely than expected from conventional normal statistics. Furthermore, the autocorrelation function peaks at multiples of a quarter-hour, the  smallest interval of electricity trading in Europe, which are most strongly pronounced after one hour (Fig.~\ref{fig:scenarios}d). All these characteristic patterns were well reproduced by our PIML model. 

Moreover, the daily pattern strongly determines the grid frequency in Continental Europe \cite{krusePredictabilityPowerGrid2020}. Our synthetic time series adequately reproduced this important pattern, especially during the evening and night (Fig.~\ref{fig:scenarios}e,f). In particular, the samples captured the specific dynamics of August and September 2019 in contrast to the daily profile predictor, which predicts a strong downward ramp at 20:00. Upward frequency deviations, for example at 05:00 and 06:00 in the morning, are reproduced less accurately than negative deviations at 23:00 (Fig.~\ref{fig:scenarios}g). Our model does not capture continuous (conventional) generation ramps that start before the beginning of the interval (cf.~Sec.~\ref{sec:performance}). These effects were particularly strong for upward ramps (e.g., at 05:50 in Fig.~\ref{fig:scenarios}g) and less dominant for downward ramps (e.g., at 23:50 in Fig.~\ref{fig:scenarios}e), which explains the model deviations in the morning interval. 

Furthermore, our results reveal important aspects of load-frequency dynamics and control. The success of our model suggests that the non-normal statistics is a direct consequence of the non-autonomous character of the power system. The changing system parameters induce heavy tails in the frequency distribution, without the need for heavy-tailed power fluctuations, cf.~the discussion in \cite{ schaferNonGaussianPowerGrid2018,kraljic2022towards}.

\section{Discussion}
\label{sec:discussion}

We have developed a model of power system operation that integrates both the internal system dynamics and the external techno-economic features. The integration has been achieved by the combination of an explicit simulation model in terms of stochastic differential equations, and an artificial neural network to link the external influences to the system parameters. We thus obtained a generic physics-inspired machine learning model of power system dynamics and control.

Using grid frequency recordings from the Continental European power grid as a test case, we demonstrated three applications of our physics-inspired model. First, we provided a probabilistic prediction of the grid frequency in intervals of 15 minutes. Our model outperformed the daily average profile of the grid frequency, which already is a good predictor in Continental Europe \cite{kruse2021revealing}. This was also possible when using only day-ahead available techno-economic features, thus providing the possibility to forecast grid frequency dynamics 15 minutes ahead. Previous grid frequency predictors only used historic frequency data as inputs \cite{krusePredictabilityPowerGrid2020, bangForecastingElectricNetwork2019, dongFrequencyPredictionPower2014}. Approaches that integrated external features previously focused on aggregated frequency deviations \cite{kaurPowerGridFrequency2013, kruse2021revealing}, which is also a cause of the data quality. Techno-economic features are typically available only on aggregated time scales of 15 min or 1h, while the frequency fluctuates on time scale of seconds (and even shorter time scales). We bridge the gap between large-scale techno-economic features and short-term frequency dynamics by using a physics-inspired model. It connects the time-aggregated features with dynamical parameters of a stochastic process that well describes the short term grid frequency fluctuations.  

Second, our model provides a tool for system identification and explanation. The model inferred the effective system parameters for every 15 minute interval from frequency measurements and techno-economic input features. The parameters were rescaled by the inertia (cf.~Sec.~\ref{sec:methods}), but the actual system parameters can be obtained by incorporating inertia time series, which however can only be approximated for large-scale power systems \cite{entso-InertiaReport, homanGridFrequencyVolatility2021}. As an example, we extracted the time-varying nature of deterministic power imbalances, which arises due to the step-wise evolution of scheduled conventional generation. The inferred power steps were particularly large for the first 15 minutes within the hour, which is consistent with the large share of generation traded at hourly day-ahead markets \cite{IntradayTradingIncreases}. The strong time variation of the inferred parameters indicates the importance of modelling the grid frequency as a non-autonomous system with time-dependent parameters. Explaining the inferred parameters with SHAP values further revealed their dependency on techno-economic drivers. For example, the primary control time scale increased with rising feed-in of wind power, which are harder to control and thus effectively lead to longer relaxation times. Our tool therefore extracts and explains physically meaningful system parameters and their time-dependent drivers.  

Third, we used our model for scenario generation of synthetic grid frequency time series. The synthetic data well approximated the heavy-tailed distribution of frequency deviations and the recurrent patterns in its autocorrelation. In contrast to previous stochastic models \cite{kraljic2022towards, vorobevDeadbandsDroopInertia2019, gorjaoDatadrivenModelPowergrid2019}, the synthetic time series also reproduced the actual frequency trajectory with its local time-dependent characteristics. Most interestingly, we only applied Gaussian white noise, but well reproduced the heavy-tailed distribution due to the time-dependent parameters of our stochastic model of power imbalance fluctuations and the control system. Notably, the model requires very little system specific information as inputs, but learns them directly from the data. Hence, the model is highly flexible and can easily be transferred to other grids. 

In the context of power system dynamics and control, physics-inspired machine learning methods have become popular during the past years \cite{chatzivasileiadisMachineLearningPower2022,misyrisPhysicsInformedNeuralNetworks2020a}. Classical physics-informed neural networks (PINNs) are commonly applied to the differential equations directly \cite{raissiPhysicsinformedNeuralNetworks2019}, which we circumvented by solving our system analytically. However, an analytical solution is not possible anymore when including non-linearities such as deadbands. In the future, our model can be modified to leverage classical PINNs to also treat non-linearities and more generic power system dynamics. Previous applications of PINNs to power system dynamics have successfully addressed autonomous dynamics \cite{misyrisPhysicsInformedNeuralNetworks2020a, huangApplicationsPhysicsInformedNeural2022}. We contribute to these developments by proposing a model that explicitly models non-autonomous dynamics, which may greatly advance the application of physics-inspired machine learning in the energy sector.

\section*{Acknowledgements}
This work was performed as part of the Helmholtz School for Data Science in Life, Earth and Energy (HDS-LEE) and received funding from the Helmholtz Association of German Research Centres via the grant no. VH-NG-1727. We gratefully acknowledge support from the German Federal Ministry of Education and Research (BMBF grant no. 03EK3055B). This project has received funding from the European Union’s Horizon 2020 research and innovation programme under the Marie Sk\l{}odowska-Curie grant agreement No. 840825. 

\bibliography{bibliography.bib}

\clearpage

\onecolumngrid

\appendix

\section{A stochastic model of grid frequency dynamics and control}
\label{app:freq-model}

\subsection{The aggregated swing equation}
\label{app:aggregate_swing_eq}

Grid frequency deviations $\Delta f(t) = (f(t) - f_{\rm ref})$ from the reference $f_{\rm ref}=$ 50 Hz or 60 Hz reflect power imbalances $\Delta P$ in the grid, which have to be compensated via different sources. Firstly, the rotational energy of synchronous machines provides momentary reserve power $P_{\rm rot}$ through changes in the rotation speed \cite{ulbigImpactLowRotational2014}:
\begin{align}
    P_{\rm rot} = \frac{2HS_B}{f_{\rm ref}} \ddt{\Delta f} = 2 \pi M \ddt{\Delta f} \, .
\end{align}
Here, $H$ denotes the average inertia constant of a synchronous machine, which is typically around $6$ s for conventional generators \cite{vorobevDeadbandsDroopInertia2019} and $S_B$ is the total rated power of all generators. Loosely speeking, $H$ equals the kinetic energy of the rotating machine rotating at $f_{\rm ref}$ divided by its rated power. The parameter $M$ then denotes the aggregated inertia of the grid.

Secondly, damping power is provided via primary control, also refered to as frequency containment reserve (FCR), and frequency-sensitive loads  \cite{ulbigImpactLowRotational2014},
\begin{align}
    P_{\rm prim} = K_1\Delta f, 
    \label{eq:app-pri-control}
\end{align}
where $K_1 = K_T + K_L$ is the inverse droop coefficient, which comprises the effect of control $K_T$ and load damping $K_L$ ($[K_1]=$ W/Hz). The control effect $K_T$ is typically one or two orders of magnitude larger than $K_L$ \cite{machowskiPowerSystemDynamics2008}. Their values are often provided in the \textit{per unit} (pu) system with $K_1^{pu}= f_{\rm ref} K_1/P_0$, as the parameter $K_T$ depends on the steady-state load $P_0$ within the specific system. For example, in Great Britain a typical value of  $K_1^{pu}=12.5$ is reported \cite{vorobevDeadbandsDroopInertia2019}. 

Thirdly, secondary control, also referred to as frequency restoration reserve (FRR), restores the frequency back to its reference $f_{\rm ref}$. Secondary control is typically implemented as an integral controller (but other implementations exist)\cite{weitenbergRobustDecentralizedSecondary2019}:
\begin{align}
    P_{\rm sec} = K_2  \bar \theta, 
        \label{eq:app-sec-control}
\end{align}
with the integrated frequency deviation
\begin{align}
    \bar \theta(t) = \int^t_{t_i} \, \Delta f(t^{\prime}) \, \d t^{\prime} .
\end{align}

The parameter $K_2$ is the secondary control gain ($[K_2]=$ W), which reads $K_2^{pu}=f_{\rm ref} K_2 / P_0$ in the pu system. For example, in Great Britain a typical value of $K_2^{pu}=0.05 / $s is reported \cite{vorobevDeadbandsDroopInertia2019}. In interconnected power grids, secondary control may also be used to reduce unscheduled flows between different control areas \cite{bottcher2020time}. These control actions are applied reciprocally in two areas and thus have only minor effects for the overall frequency dynamics.

Finally, this yields the aggregated swing equation, which is simply the power balance of all contributions:
\begin{align}
    \ddt{\bar \theta} &= \Delta f ,\\
    2\pi M \ddt{\Delta f} &= - K_1 \Delta f - K_2 \bar \theta + \Delta P(t).
    \label{eq:swing_eq}
\end{align}
Here, $\Delta P(t)$ denotes the imbalance of power generation and load, excluding the contribution of the control system \eqref{eq:app-pri-control} and \eqref{eq:app-sec-control}.

\subsection{Stochastic differential equations}
\label{app:sde}

Following reference \cite{gorjaoDatadrivenModelPowergrid2019}, we modelled the power imbalances $\Delta P(t) =  P_{\rm im}(t) + \bar D \xi(t) $ as a sum of ``deterministic'' power imbalances $P_{\rm im}(t)$ and stochastic deviations $\bar D \xi(t)$. In the large European power grids, deterministic power imbalances repeatedly arise due to a different ramping of dispatchable generators and the load or due to forecasting errors \cite{hirthBalancingPowerVariable2015}. Stochastic deviations are modelled as Gaussian noise $\xi(t)$ defined as the derivative of a Wiener process $W_t$ that has independent, normally distributed increments $\d W_t$ with zero mean and variance $\mean{\d W_t^2} = \d t$. 

To estimate model parameters, we have to rescale Eq.~\eqref{eq:swing_eq} by $M$ as all parameters otherwise are undefined up to a multiplicative factor. In addition, we transition from frequencies and integrated frequencies to angular velocities and angles. We define
\begin{align}
    \tau^{-1} &= \frac{K_1}{2\pi M}, \\
    \kappa^{-2} &= \frac{K_2}{2\pi M}, \\
    D &= \frac{\bar D}{M}, \\
    P(t) &= \frac{P_{\rm im}(t)}{M}, \\
    \theta &= 2\pi \bar \theta .
\end{align}
Based on these definitions, we obtain a stochastic model for the angular grid frequency deviation $\omega = 2\pi \Delta f$: 
\begin{align}
    \d \theta &= \omega \, \d t , \label{eq:theta_eq_app}\\
    \d \omega &= \left( P(t) - \tau^{-1} \omega - \kappa^{-2} \theta \right) \d t + D \,  \d W_t . \label{eq:omega_eq_app}
\end{align}
Collecting both stochastic variables into a vector $\v X = (\theta, \omega)^T$, we can write our model as a two-dimensional matrix equation
\begin{align}
    \d \v X = \v a(\v X, t) \d t + \v D \, \d \v W_t, \label{eq:matrix_sde}
\end{align}
using the drift vector $\v a (\v X, t) = (a_\theta, a_{\omega})^T$ with $a_\theta = \omega$, 
$a_{\omega}=P(t) -\tau^{-1} \omega- \kappa^{-2} \theta$, the diffusion matrix $\v D = \textrm{diag}(D,0)$ and a two-dimensional Wiener process $\v W_t$.

The stochastic differential equation \eqref{eq:matrix_sde} can be recast into a Fokker-Planck equation for the joint probability density function $\mathcal{P}(\theta, \omega; t)$ that describes the distribution of the two random variables $\theta$ and $\omega$ at time $t$ \cite{gardiner2009handbook}. Using It{\^o} calculus, one obtains
\begin{align}
    \frac{\partial}{\partial t} \mathcal{P}(\theta, \omega; t) 
    &= \bigg[ -\frac{\partial}{\partial \omega}
    \left(  P(t) - \tau^{-1} \omega - \kappa^{-2} \theta    \right)
     - \frac{\partial}{\partial \theta} \omega
    + \frac{D^2}{2} \frac{\partial^2}{\partial \omega^2}
    \bigg]
    \mathcal{P}(\theta, \omega; t)    .
    \label{eq:app-fpe}
\end{align}

\subsection{Solution of the Fokker-Planck equation}
\label{app:FPE-sol}

In this section, we proof that the Fokker-Planck equation \eqref{eq:app-fpe} is solved by a multivariate normal distribution with PDF
\begin{align}
    \mathcal{P}(\v x; t) 
    = \frac{1}{2\pi | \Sigma |} 
    \exp\left(  -\frac{1}{2} (\v x -\v \mu)^\top 
    \Sigma^{-1} (\v x -\v \mu) \right)
    \label{eq:app-pdf-gauss}
\end{align}
with $\v x^\top = (\theta,\omega)$ and time-dependent parameters
\begin{align*}
    \v \mu(t) = \begin{pmatrix}
    \mu_\theta(t) \\ \mu_\omega(t)
    \end{pmatrix},
    \quad
    \Sigma(t) = \begin{pmatrix}
    \sigma^2_\theta(t) & \sigma_{\theta \omega}(t) \\
    \sigma_{\theta \omega}(t) & \sigma^2_\omega(t) 
    \end{pmatrix} \, .
\end{align*}
if the parameters satisfy the ordinary differential equations
\begin{align}
    \frac{d}{dt} \mu_\theta &= \mu_\omega \, ,\nonumber \\
    \frac{d}{dt} \mu_\omega &= P(t) - \tau^{-1} \mu_\omega - \kappa^{-2} \mu_\theta\, , \nonumber \\
    \frac{d}{dt} \sigma^2_\theta &= 2 \sigma_{\theta \omega}, \nonumber \\
    \frac{d}{dt} \sigma^2_\omega &= \sigma^2\omega - \tau^{-1}  \sigma_{\theta \omega} - \kappa^{-2} \sigma^2_\theta\, , \nonumber \\
    \frac{d}{dt} \sigma_{\theta \omega} &= -2 \tau^{-1} \sigma^2_\omega -2 \kappa^{-2} \sigma_{\theta \omega} \, .
    \label{eq:app-eom-musigma}
\end{align}

We proof this result using the characteristic function, which is defined via the Fourier transform
\begin{align}
    \phi(\v u;t) = \int e^{i \v u^\top \v x} \,
      \mathcal{P}(\v x;t) d^2 \v x.
\end{align}
In terms of the characteristic function, the FPE reads
\begin{align}
    \frac{\partial}{\partial t} \phi(\v u;t)
    &= \mathcal{L}  \phi(\v u;t) \nonumber \\
    &= is \bigg[ \left( P(t) +i \tau^{-1} \frac{\partial}{\partial s} +i \kappa^{-2} \frac{\partial}{\partial r} \right)   + r \frac{\partial }{\partial s} 
    - \frac{D^2}{2} s^2 \bigg]  \phi(\v u;t), 
    \label{eq:app-fpe-cf}
\end{align}
where we have defined $\v u^\top = (r,s)$. The characteristic function of the normal distribution \eqref{eq:app-pdf-gauss} reads
\begin{align}
        \label{eq:app-cf-gauss}
    \phi(\v u;t) &= \exp\left( i \v u^\top \v \mu
    - \frac{1}{2} \v u^\top \Sigma \v u \right) \, , \\
    \phi(r,s;t) &=
    \exp \bigg( i (r \mu_\theta + s \mu_\omega)  -\frac{1}{2} (r^2 \sigma^2_\theta + s^2 \sigma_\omega^2 + 2 rs \sigma_{\theta \omega})   \bigg).
    \nonumber
\end{align}

We now show that the normal distribution \eqref{eq:app-cf-gauss} with the parameters
\eqref{eq:app-eom-musigma} satisfies the Fokker-Planck equation \eqref{eq:app-fpe-cf}. We first evaluate the right-hand side of the FPE,
\begin{align}
    \mathcal{L}  \phi(\v u;t)
    &= is \left[ \left( P(t) +i \tau^{-1} \frac{\partial}{\partial s} +i \kappa^{-2} \frac{\partial}{\partial r} \right)
    + r \frac{\partial }{\partial s} 
    - \frac{D^2}{2} s^2 \right]   \exp \left( i (r \mu_\theta + s \mu_\omega)
    -\frac{1}{2} (r^2 \sigma^2_\theta + s^2 \sigma_\omega^2 + 2 rs \sigma_{\theta \omega})   \right), \nonumber \\
    &= \left[ i s P(t)- \frac{D^2}{2} s^2
    - (\tau^{-1} s -r )
    (i \mu_\omega - s \sigma^2_\omega - r \sigma_{\theta \omega} )
    - \kappa^2 s (i \mu_\theta - r \sigma^2_\theta - s \sigma_{\theta \omega})
    \right] \phi(\v u;t).
    \label{app:proof-normal-rhs}
\end{align}
Now we proceed with the left-hand side,
\begin{align}
   \frac{\partial}{\partial t} \phi(\v u;t)
   &= \left[
   i \left( s \frac{d \mu_\omega}{dt} + r \frac{s \mu_\theta}{dt} \right) 
   - \frac{1}{2} \left(
   r^2 \frac{d \sigma^2_\theta}{dt}
   + s^2 \frac{d \sigma^2_\theta}{dt}
   + 2rs \frac{d \sigma_{\theta \omega}}{dt}
    \right) \right] \phi(\v u;t) \, .
\end{align}  
Inserting the equations \eqref{eq:app-eom-musigma} then yields
\begin{align}
    \frac{\partial}{\partial t} \phi(\v u;t)
    &= \left[ i s P(t) - \frac{D^2}{2} s^2
    - (\tau^{-1} s -r )
    (i \mu_\omega - s \sigma^2_\omega - r \sigma_{\theta \omega} )
    - \kappa^2 s (i \mu_\theta - r \sigma^2_\theta - s \sigma_{\theta \omega})
    \right] \phi(\v u;t),
\end{align}
which coincides with the right-hand side (\ref{app:proof-normal-rhs}).

\subsection{Moment equations}

The ordinary differential equations for the parameters \eqref{eq:app-eom-musigma} can also be obtained in a more direct way, once we know that the PDF remains Gaussian for all times. In fact, we can exploit that the parameters of a Gaussian PDF equal the mean and the (co-) variances. The dynamics of the mean and the (co-) variances are determined by the moment equations, which we extracted using It\^o's lemma. For any twice differentiable scalar function $g(\v X)$ of the random variable $\v X$ in Eq.~\eqref{eq:matrix_sde}, It\^o's lemma reads \cite{oksendalStochasticDifferentialEquations2003},
\begin{align}
    \d g = \left[ \left(\v \nabla g\right)^T\v \mu + \frac{1}{2} \rm{Tr} \left( \v{D}^T \v{H}_g \v{D}\right) \right] \d t +   \left(\v \nabla g\right)^T\v D \d \v W,
\end{align} 
where $\v \nabla g$ is the gradient and $\v H_g$ is the hessian matrix of the function $g(\v X)$. This yielded in our case
\begin{align}
    \d g = \left[ \frac{\partial g}{\partial \omega} \mu_{\omega} + \frac{\partial g}{\partial \theta}\omega + \frac{D^2}{2}\frac{\partial^2 g}{\partial \omega^2} \right] \d t + \frac{\partial g}{\partial \omega} D \, \d W.
\end{align}
To apply this to moment functions, we further assumed $\d \mean g  = \mean{ \d g }$. For the first moments (averages) $g=\mean \theta $ and $g=\mean \omega $ we obtained
\begin{align}
   \ddt{\mean \theta}&= \mean \omega,  \\
    \ddt{\mean \omega} &=  P(t) - \tau^{-1} \mean \omega - \kappa^{-2} \mean\theta.
\end{align}
The second moments  $g=\mean{\theta^2}$, $g= \mean{\omega^2}$ and the mixed moment $g=\mean{\omega\theta}$ yielded
\begin{align}
    \ddt{\mean{\theta^2}} &= 2\mean{\theta\omega}, \\
    \ddt{\mean{\omega^2}} &= 2P(t) \mean \omega - 2\tau^{-1} \mean{\omega^2} - 2 \kappa^{-2} \mean{\omega\theta} + D^2, \\
    \ddt{\mean{\omega\theta}} &= P(t) \mean \theta - \tau^{-1} \mean{\omega\theta}-\kappa^{-2} \mean{\theta^2} + \mean{\omega^2} .
\end{align}
In this derivation, we used $\mean{\omega\d W}=0$ and $\mean{\theta \d W}=0$. 
Identifying 
$\mu_\theta = \mean \theta$, 
$\mu_\omega = \mean \omega$,
$\sigma_{\omega}^2 = \mean{\omega^2}-\mean{\omega}^2$, 
$\sigma_{\theta}^2 = \mean{\theta^2}-\mean{\theta}^2$ and 
$\sigma_{\theta, \omega} = \mean{\omega\theta} - \mean \omega \mean \theta$
then reproduces Eq.~\eqref{eq:app-eom-musigma}.

\subsection{Solution of the moment equations}
\label{app:sol_moment_eq}

We now provide a semi-analytic solutions for the ordinary differential equations \eqref{eq:app-eom-musigma} describing the evolution of the paramaters
$\mu_\theta$, $\mu_\omega$, $\sigma_{\omega}^2$, $\sigma_{\theta}^2$ and $\sigma_{\theta, \omega}$. We first note that the equations for the deterministic part (the means) and the stochastic part (the (co-) variances) decouple, hence they can be treated separately.  We collected the deterministic equations using the vector $\v y_d = (\mu_\theta, \mu_\omega)^T$,
\begin{align}
    \ddt{\v y_d} &= \v A_d \v y_d + \v b_d(t), \\
    \v A_d &= \begin{pmatrix} 0 & 1 \\ -\kappa^{-2} & -\tau^{-1} \end{pmatrix}, \\
    \v b_d(t) &= \begin{pmatrix} 0 \\ P(t) \end{pmatrix}.
\end{align}
With $\v y_s=(\sigma_{\theta}^2, \sigma_{\theta, \omega}, \sigma_{\omega}^2)^T$, the stochastic part yielded
\begin{align}
    \ddt{\v y_s} &= \v A_s \v y_s + \v b_s, \\
    \v A_s &= \begin{pmatrix} 0 & 2 & 0 \\ -\kappa^{-2} & -\tau^{-1} & 1 \\ 0 & -2\kappa^{-2} & -2\tau^{-1} \end{pmatrix}, \\
    \v b_d &= \begin{pmatrix} 0 \\ 0 \\ D^2 \end{pmatrix}.
\end{align}
These equations are linear, ordinary differential equations (ODEs), for which several solution methods exist.

\subsubsection{Solution of the homogeneous equations}

The general solution of the homogeneous ODE $\dot{\v y} = \v A \v y $ with time-independent coefficients $\v A$ is given by
\begin{align}
    \v y_h(t) = \v U(t)\v{U}^{-1}(0)\v{y}_0.
    \label{eq:general_hom_sol}
\end{align}
The columns of the matrix $\v U(t)$ span the solution space of the homogeneous ODE. The column vectors are given by $\v u_i = \v v_i e^{\lambda_i t}$, where $\v v_i$ are the eigenvectors of $\v A$ and $\lambda_i$ are the corresponding eigenvalues. Using any computer algebra program, we calculated the eigenvalues and eigenvectors for the matrix $\v A_d$ of the deterministic part:
\begin{align}
    \lambda_{d,1} &= -\frac{1}{2\tau}\sqrt{1-\frac{4\tau^2}{\kappa^2}} - \frac{1}{2\tau} ,\\
    \lambda_{d,1} &= \frac{1}{2\tau}\sqrt{1-\frac{4\tau^2}{\kappa^2}} - \frac{1}{2\tau} ,\\
    \v v_{d,1} &= (\lambda_{d,2}\kappa^2 ,1 )^T ,\\
    \v v_{d,2} &= (\lambda_{d,1}\kappa^2 ,1 )^T .
\end{align}
For the matrix $\v A_s$ from the stochastic part, we obtained
\begin{align}
    \lambda_{s,1} &= -\tau^{-1},\\
    \lambda_{s,2} &= -\frac{1}{\tau}\sqrt{1-\frac{4\tau^2}{\kappa^2}} - \frac{1}{\tau} ,\\
    \lambda_{s,3} &= \frac{1}{\tau}\sqrt{1-\frac{4\tau^2}{\kappa^2}} - \frac{1}{\tau} ,\\
    \v v_{s,1} &= (\kappa^2, -\frac{\kappa^2}{2\tau}, 1)^T ,\\
    \v v_{s,2} &= (-\kappa^2 - \lambda_{s,3}\frac{\kappa^4}{2\tau}, \frac{\lambda_{s,3} \kappa^2}{2}, 1)^T ,\\
    \v v_{s,3} &= (-\kappa^2 - \lambda_{s,2}\frac{\kappa^4}{2\tau}, \frac{\lambda_{s,2} \kappa^2}{2}, 1)^T  .
\end{align}

To specify the homogeneous ODE solutions for $\mu_\omega(t)$ and $\sigma_{\omega}^2(t)$, we needed the inverses of the matrix $\v U(t)$, i.e., the matrices $\v{U}_d^{-1}(t)$ and $\v{U}_s^{-1}(t)$, which we again obtained through a computer algebra program.
The $\omega$-components of the solution for the homogeneous system \eqref{eq:general_hom_sol} then read
\begin{align}
    \mu_{\omega,h}(t) &= \frac{1}{\lambda_{d,1}-\lambda_{d,2}}
    \left[ \kappa^{-2} \theta_0 \left(e^{\lambda_{d,2}t}-e^{\lambda_{d,1}t} \right)  + \omega_0 \left(\lambda_{d,1}e^{\lambda_{d,1}t}-\lambda_{d,2}e^{\lambda_{d,2}t} \right) \right] ,\\
    \sigma_{\omega,h}^2(t) &= \left[ \sigma_{\theta,0}^2 \left(e^{\lambda_{s,2}t}+e^{\lambda_{s,3}t} -2e^{\lambda_{s,1}t} \right)\frac{\tau^2}{\kappa^2} \right. \\ \nonumber
    +& \sigma_{\omega, \theta, 0} \left(-2\tau e^{\lambda_{s,1}t} + \frac{8\tau^2}{\kappa^2} \frac{\lambda_{s,2}\kappa^2(4\tau)^{-1} +1}{\lambda_{s,2} -\lambda_{s,3}}e^{\lambda_{s,2}t} +\frac{8\tau^2}{\kappa^2} \frac{\lambda_{s,3}\kappa^2(4\tau)^{-1} +1}{\lambda_{s,3} -\lambda_{s,2}}e^{\lambda_{s,3}t} \right) \\
    +&\left.  \sigma_{\omega, 0}^2 \left(-2\tau e^{\lambda_{s,1}t} + 
    \frac{2\lambda_{s,2}\tau^2-\lambda_{s,2}\kappa^2-2\tau}{\lambda_{s,3} -\lambda_{s,2}}e^{\lambda_{s,2}t} +\frac{2\lambda_{s,3}\tau^2-\lambda_{s,3}\kappa^2-2\tau}{\lambda_{s,2} -\lambda_{s,3}}e^{\lambda_{s,3}t} \right) 
     \right] \frac{1}{\kappa^2-4\tau^2},
\end{align}
where $\theta_0, \omega_0$ denote the initial conditions of the averages and $\sigma_{\theta,0}^2,\sigma_{\theta,\omega,0},\sigma_{\omega,0}^2$ represent the initial conditions of the covariances.

\subsubsection{(Semi-)Analytical solution of inhomogeneous equations}

The general solution of the inhomogeneous ODE $\dot{\v y} = \v A \v y + \v b(t)$ with time-independent coefficient $\v A$ is given by the sum of the homogeneous solution and an inhomogeneous contribution $\v y_{in}(t)$ 
\begin{align}
    \v y_{\textrm{\tiny ODE}}(t) &= \v y_h(t) + \v y_{in}(t) \nonumber,\\
    &= \v y_h(t) + \v{U}(t) \int_{0}^{t} \v{U}^{-1}(t')\v b(t') \d t'.
\end{align}
We first provide a semi-analytical solution, which leaves the integration of the inhomogeneity $b_d(t)$ to a numerical routine. This enables us to flexibly insert different power function $P(t)$, as we will see below.

The $\omega$-components of the inhomogeneous contributions yield
\begin{align}
    \mu_{\omega,in}(t) &=  \frac{1}{\lambda_{d,1}-\lambda_{d,2}}
    \left[ \lambda_{d,1}e^{\lambda_{d,1}t} \int_0^t P(t') e^{-\lambda_{d,1}t'} \d t'  - \lambda_{d,2}e^{\lambda_{d,2}t} \int_0^t P(t') e^{-\lambda_{d,2}t'} \d t' \right] ,
    \label{eq:omega_in_sol}\\
    \sigma_{\omega, in}^2(t) &= \frac{D^2}{\kappa^2-4\tau^2} \left[ \frac{2\tau^2}{\lambda_{s,1}}(1-e^{\lambda_{s,1}t})+ \frac{2\tau/\lambda_{s,2} -2\tau^2+\kappa^2}{\lambda_{s,3}-\lambda_{s,2}} (1-e^{\lambda_{s,2}t}) + \frac{2\tau/\lambda_{s,3} -2\tau^2+\kappa^2}{\lambda_{s,2}-\lambda_{s,3}} (1-e^{\lambda_{s,3}t}) \right].
    \label{eq:sigma_in_sol}
\end{align}
Note that the deterministic part contains an integral over the deterministic power imbalance $P(t)$, which we can compute numerically for any power function. However, the factors $e^{-\lambda t}$ in Eq.~\eqref{eq:omega_in_sol} can become very large as $\lambda<0$ for stable systems thus causing numerical problems. To use this semi-analytical solution during neural network training, one has to mitigate these numerical problems, e.g., by strongly restricting the parameter space. 

We used a fully analytical solution for the case $P(t)=q+rt$, thus avoiding these numerical problems. In this case, the inhomogeneous solution of the deterministic part yielded
\begin{align}
     \mu_{\omega,in}(t) = \frac{1}{\lambda_{d,1}-\lambda_{d,2}} \left[e^{\lambda_{d,1}t}\left( q+\frac{r}{\lambda_{d,1}}\right) - e^{\lambda_{d,2}t}\left( q+\frac{r}{\lambda_{d,2}}\right) + r \left(\frac{1}{\lambda_{d,2}} - \frac{1}{\lambda_{d,1}}\right) \right].
\end{align}

Note that we only require the marginal probability density $\mathcal P(\omega;t) = \int \mathcal P(\theta, \omega;t) \, \textrm{d} \theta$ to model the grid frequency dynamics, hence we only needed a closed form solution for $\mu_{\omega}$ and $\sigma_{\omega}$.

\section{Parameter values from the literature}
\label{app:param-ref-values}

Based on the definition in Eqs.~\eqref{eq:theta_eq_app}-\eqref{eq:omega_eq_app}, Gorjao et al.~inferred time-independent parameter values using the Kramers-Moyal expansion \cite{gorjaoDatadrivenModelPowergrid2019}. The authors employed the actual grid frequency instead of the angular velocity. Thus, we rescaled the results according to $D \rightarrow 2\pi D $, $q \rightarrow 2\pi q$, $r\rightarrow 2\pi r$, $\theta\rightarrow 2\pi \theta$, while $\tau$ and $\kappa$ stayed the same.
\begin{align}
    \tau &= 120~s \, ,\\
    \kappa &= 183~s \, ,\\
    D &= 0.007~s^{-3/2} \, , \\
    q &= 0.0042~s^{-2} \, ,\\
    r &= 0.000009~s^{-3} .
\end{align}
In the main text, we use these parameter values as reference values and therefore call them $\vartheta_j^{\rm ref}$. 
Note that the authors of ref. \cite{gorjaoDatadrivenModelPowergrid2019} did not directly provide a result for $r$, but its value was implicitly defined through the constraint $\langle P_{\rm im}(t) \rangle_t = 0$, which yields $r=2q/t_{max}$. Moreover, the parameter $q$ was specified separately for the full hour and for every (other) quarter of the hour, so we took the average value here.

\section{Parameter Scaling and Constraints}
\label{app:scalings}

\begin{table*}[tb]
    \caption{Properties of dynamical system parameters that are predicted by the FFNN (cf.~Fig.~\ref{fig:method}d). The parameters are subject to several physical constraints which are summarised in the third row. The output of the FFNN is rescaled by constant factors listed in the fourth row to improve training efficiency (referred to as the standard scaling). To test the impact of the scaling, we varied the scaling $s_j$ according to the parameter choices in the fifth row. Furthermore, we ensure that the dynamical system parameter exceeds a minimum value listed in the sixth row. }
    \label{tab:constraints}
    \centering
    \begin{tabular}{c|c|c|c|c|c|c|c|c}
         Parameter $\vartheta_j$ & $\vartheta_1$ & $\vartheta_2$ & $\vartheta_3$ & $\vartheta_4$ & $\vartheta_5$ & $\vartheta_6$ & $\vartheta_7$ & $\vartheta_8$  \\ \hline 
         Name & $\sigma_{\theta,0}$ & $\sigma_{\theta, \omega, 0}$ & $\sigma_{\omega,0}$ & $\tau$ & $\kappa$ & $D$ & $q$ & $r$ \\ \hline
         Physical constraints & $\sigma_{\theta,0}>0$ & $|\sigma_{\theta, \omega, 0}|\le\sigma_{\theta,0}\sigma_{\omega,0}$ & $\sigma_{\omega,0}>0$ & $0<\tau\le\kappa/2$ & $\kappa>0$ & $D>0$ & $r \in \mathbb{R}$ & $q \in \mathbb{R}$ \\ \hline  
         Scaling $s_j$ & 0.01 & - & 0.1 & - & 100 & 0.01 & $10^{-3}$ & $10^{-6}$  \\ \hline
         Scaling variation & $\{1 \}$ & - & $\{ 0.1\}$ & - & $\{1000,100\}$ & $\{ 0.01, 0.1\}$ & $\{10^{-3}, 10^{-2} \}$ & $\{10^{-5}, 10^{-6} \}$ \\ \hline
         Minimum $v_j$ & $10^{-3}$ & - & $10^{-3}$ &10 & 30 &$10^{-4}$ & - & - \\ \hline
         Range of $\nu_j(u_j)$ &  $(v_1, \infty)$ & $(-\sigma_{\theta,0}\sigma_{\omega,0}, \sigma_{\theta,0}\sigma_{\omega,0})$ & $(v_3, \infty)$ & $(v_4, \kappa/2)$ & $(v_5, \infty)$ & $(v_6, \infty)$ & $(-\infty, \infty)$ & $(-\infty, \infty)$ 
    \end{tabular}
\end{table*}

The developed PIML model includes a layer that rescales the parameters and ensures some physical constrains (Fig.~\ref{fig:method}d). The outputs $u_j$ of the FFNN do not necessarily fulfil the physical constraints of the parameters $\vartheta_j$ (cf.~Tab.~\ref{tab:constraints}), as the linear activation of the output takes arbitrary real values, while $\tau$, for example, only takes positive real values. Moreover, the physical parameters $\vartheta_j$ vary strongly in scale (cf.~appendix \ref{app:param-ref-values}), but the outputs $u_j$ of the initialised FFNN typically have the same scale due uniform random initialisation of the weights \cite{hastieElementsStatisticalLearning2016}. This will yield large initial errors along certain parameter axis thus leading to inhomogeneous loss landscapes which can make optimisation inefficient and more difficult \cite{mehtaHighbiasLowvarianceIntroduction2019}. 

Therefore, we added a constraint and scaling layer that applies functions $\nu_j$ to the FFNN outputs. The results then represent the parameter estimates $\vartheta_{j} = \nu_j(u_j)$. First, the functions $\nu_j$ enforce the physical constraints. For example, a softplus function $\sp(u) = \log(\exp(u) + 1) \in (0, \infty)$ enforces positivity, and the sigmoid function $\textrm{Sig}(u) = (1+\exp(-u)^{-1} \in (0,1)$ was used to ensure that $\tau\le \kappa/2$ holds. Numerical imprecision can lead to a violation of these constraints so that we added a safety factor $\delta=0.999$ in some cases. Second, the factors $s_j$, which mirror the typical scale of parameters $\vartheta_j$, are applied to make the optimisation more efficient. Third, minimum values are added in certain cases to ensure numerical stability during optimisation. For example, a very small standard deviation $\sigma_{\omega,0}$ can lead to probability densities beyond float precision. All in all, we defined the following functions using minimum values, scaling factors and constraints from Tab.~\ref{tab:constraints}: 
\begin{align*}
    \nu_1(u_1) &= \sp(u_1)s_1 + v_1,\\
    \nu_2(u_2) &= \delta\tanh(u_1) \nu_1(u_1) \nu_3(u_3),\\
    \nu_3(u_3) &= \sp(u_3)s_3 + v_3,\\
    \nu_4(u_4) &= \left(\frac{2 v_4}{\nu_5(u_5)} + \delta \textrm{Sig}(u_4)\left(1-\frac{2 v_4}{\nu_5(u_5)} \right)\right) \frac{\nu_5(u_5)}{2},\\
    \nu_5(u_5) &= \sp(u_5)s_5 + v_5,\\
    \nu_6(u_6) &= \sp(u_6)s_6 + v_6,\\
    \nu_7(u_7) &= u_7 s_7,\\
    \nu_8(u_8) &= u_8 s_8.
\end{align*}
To test the impact of the scaling with $s_j$, we varied the scaling parameters according to Tab.~\ref{tab:constraints}. In particular, we trained the PIML model for each combination of the individual scaling choices listed in the table. For each scaling tuple, we additionally simulated 10 different random initialisation of the FFNN weights. Finally, we trained 10 initialisations using the standard scaling defined in Tab.~\ref{tab:constraints} and no scaling with $s_j=1$ (cf.~Sec.~\ref{sec:inference}).

\end{document}